\documentclass[nocopyright]{cimento}

\setcopyright{CERN on behalf the NA61/SHINE Collaboration}
\usepackage{graphicx}  % got figures? uncomment this

\title{Femtoscopy with L\'evy sources at NA61/SHINE}
\author{B.~P\'orfy\from{ins:x}\from{ins:y} for the NA61/SHINE Collaboration}
\instlist{\inst{ins:x} Wigner Research Centre for Physics, Konkoly-Thege Mikl\'os \'ut 29-33, H-1121 Budapest, Hungary
  \inst{ins:y} Department of Atomic Physics, Faculty of Science, E\"otv\"os Lor\'and University, P\'azm\'any P\'eter s\'et\'any 1/A, H-1111 Budapest, Hungary}
\begin{document}

\maketitle

\begin{abstract}
In the recent decades of high-energy physics research, it was demonstrated that strongly interacting quark-gluon plasma (sQGP) is created in ultra-relativistic nucleus-nucleus collisions. Investigation and understanding of the properties of the hadronic matter are among the most important goals of the NA61/SHINE collaboration at CERN SPS. Mapping of the phase diagram is achieved by varying the collision energy (5 GeV $\sqrt{s_{NN}}<17$ GeV) and by changing the collision system ($p$+$p$, $p$+Pb, Be+Be, Ar+Sc, Xe+La, Pb+Pb). Femtoscopic correlations reveal the space-time structure of the hadron emitting source.

In this article, we report on the measurement of femtoscopic correlations in small to intermediate systems. Comparing the measurements to calculations based on symmetric L\'evy sources, we discuss the results on L\'evy source parameters as a function of average pair transverse mass. One of the physical parameters is of particular importance, the L\'evy exponent $\alpha$, which describes the shape of the source and may be related to the critical exponent $\eta$ in the proximity of the critical point. Therefore, measuring it may shed light on the location of the critical endpoint of the QCD phase diagram.
\end{abstract}

\section{Introduction} 
\begin{figure}[h!]
\centering
\includegraphics[width=1\textwidth]{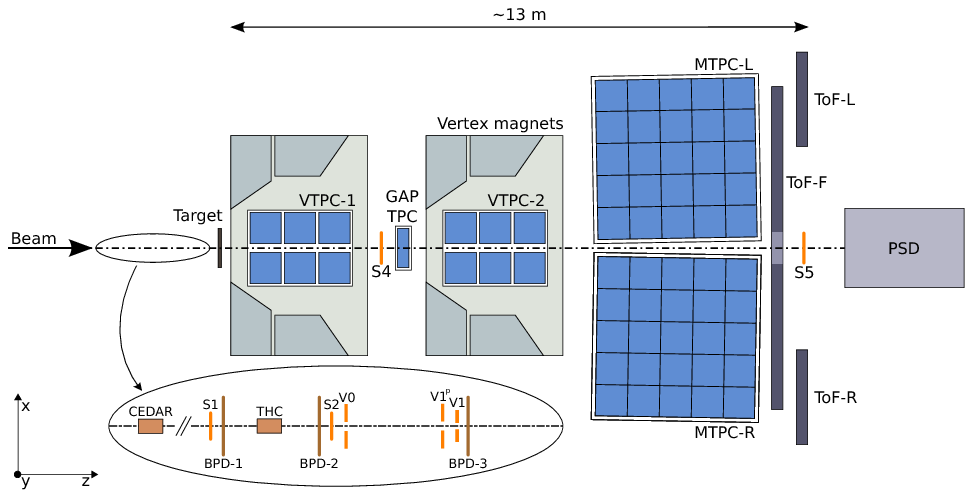}
\caption{The setup of the NA61/SHINE detector system during the run of Ar+Sc.}
\label{fig:detectorsetup}
\end{figure}

The NA61/SHINE is a fixed target experiment located in the North Area H2 beam line of the CERN SPS. The experiment houses multiple Time Projection Chambers (TPCs), covering the full forward hemisphere making it essentially a large acceptance hadron spectrometer~\cite{Abgrall:2014xwa}. The detector setup of the NA61/SHINE detector system is shown in Fig.~\ref{fig:detectorsetup}. This results in an outstanding tracking down to $p_{\rm{T}} \approx$ 0 GeV/\textit{c}. The main goals of NA61/SHINE include the investigation and mapping of the phase diagram of the strongly interacting matter at different temperatures and baryon-chemical potentials. In this proceedings, we report on our study using small ($^7$Be+$^9$Be) to intermediate ($^{40}$Ar+$^{45}$Sc) systems with 150\textit{A} GeV/\textit{c} beam momentum ( $\sqrt{s_{\rm{NN}}} = 16.82$ GeV), at 0\--20\% centrality in Be+Be and 0\--10\% centrality in Ar+Sc. The two systems are rather small with low collision energy, hence different insights may be gained here as compared to large systems. In our analysis we study the phase diagram of QCD by the method of quantum-statistical (Bose-Einstein) correlations with final state interactions using spherically symmetric L\'evy distributions, defined as: 
\begin{equation}\label{eq:levydistr}
\mathcal{L}(\alpha,R,\textbf{r})=\frac{1}{(2\pi)^3} \int \rm{d}^3\vec{\zeta} e^{i\vec{\zeta} \textbf{r}} e^{-\frac{1}{2}|\vec{\zeta} R|^{\alpha}},
\end{equation}
where $R$ is the L\'evy scale parameter, $\alpha$ is the L\'evy stability index, $\textbf{r}$ is the vector of spatial coordinates and the vector $\zeta$ represents the integration variable.  The key relationship for measuring Bose-Einstein correlations shows that the momentum correlations, $C(q)$, are related to the properties of the particle emitting source, $S(x)$, that describes the probability density of particle creation for a relative coordinate \textit{x} as $C(q) \cong 1 + | \tilde{S}(q) |^2$, where $\tilde{S}(q)$ is the Fourier transform of $S(x)$, and \textit{q} is the relative momentum of the particle pair (with the dependence on the average momentum, $K$, of the pair suppressed and described in more detail in Ref.~\cite{NA61SHINE:2023qzr}). 

The L\'evy approach allows us to describe the shape of the source. There are two special cases of the distribution defined in Eq{.}~\ref{eq:levydistr}, the Gaussian distribution for $\alpha = 2$ and Cauchy distribution with $\alpha = 1$. Unlike the Gaussian assumption, in our case a more general approach is taken by assuming the L\'evy  shape. The appearance of non-Gaussianity could be attributed to critical fluctuations and the emergence of spatial correlations on a large scale~\cite{Csorgo:2008ayr}. Furthermore, L\'evy distribution exhibits a power-law tail $\sim r^{-(1 + \alpha)}$ (in three dimensions) in the case of $\alpha < 2$, where $r \equiv |\textbf{r}|$. The critical behavior is characterized by, among others the critical exponent related to spatial correlations, $\eta$, showing a power-law structure $(-(d - 2 + \eta)$, with \textit{d} denoting dimensions. It has been suggested that the universality class of QCD is the same as that of the 3D Ising model in Refs.~\cite{Halasz:1998qr,Stephanov:1998dy}. The value of $\eta$ in the 3D Ising model is $0.03631 \pm 0.00003$~\cite{El-Showk:2014dwa}. Alternatively the universality class of the random external field 3D Ising model may be considered, corresponding to an $\eta$ value of 0.50 $\pm$ 0.05~\cite{Rieger:1995aa}. The L\'evy exponent $\alpha$ was then suggested to be directly related to or being explicitly equal to the critical exponent $\eta$~\cite{Csorgo:2005it}, in absence of other phenomena affecting the source shape. However, the appearance of L\'evy-shape of the source can be attributed to several different factors besides critical phenomena, including QCD jets, anomalous diffusion, and others~\cite{Metzler:1999zz,Csorgo:2003uv, Csorgo:2004sr, Kincses:2022eqq, Korodi:2022ohn}.  Hence, while a non-monotonic behavior of $\alpha$ is expected based on the above simple picture, for an understanding of the physical processes influencing the spatial structure of the hadron emission, measurements of $\alpha$ are needed in different collision systems at various energies. The L\'evy exponent of the source distribution can be measured using femtoscopic correlation functions. These correlation functions admit the following form for L\'evy  sources:
\begin{equation}\label{eq:corrfunc}
C^0_2(q) = 1 + \lambda \cdot e^{-(qR)^\alpha},
\end{equation}
where $C^0_2(q)$ is the correlation function without any final state interaction.
One may note that if $\alpha = 1$ then the correlation function is exponential, while if $\alpha = 2$, $C^0_2(q)$ becomes a Gaussian. The third physical parameter appearing in the above equation is the correlation strength $\lambda$ (often called intercept parameter) which is interpreted in terms of the core-halo model in Refs.~\cite{Csorgo:1999sj,Csorgo:1995bi}. The intercept parameter then can be expressed as 
\begin{equation}\label{eq:lambda}
\lambda = \left(\frac{N_{\rm{core}}}{N_{\rm{core}} + N_{\rm{halo}}}\right)^2.
\end{equation}
The core-halo model assumes that the source \textit{S} is made up of two parts, the core and the halo $S_{\rm{core}}$ and $S_{\rm{halo}}$, respectively. The core contains pions created close to the center, directly from hadronic freeze-out (primordial pions) or from extremely short lived (strongly decaying) resonances. The halo consists of pions created from longer-lived resonances and the general background. It may extend to thousands of femtometers, while core part has a size of around a few femtometers.In the above equation, \textit{N} denotes their respective multiplicities in the given event class.

\section{Measurement details}
We discuss the measurements of one dimensional two-pion femtoscopy correlation functions for identified pion pairs in Be+Be and in Ar+Sc collisions at 150A GeV/c with 0-20\% and 0–10\% centrality. We investigate the combination of positive pion pairs ($\pi^+\pi^+$) and negative pion pairs ($\pi^-\pi^-$). Event and track quality cuts were applied before particle identification (PID). The PID was done by measuring the energy loss in the TPC gas (d$E$/d$x$) and comparison with the Bethe-Bloch curves. 

The centrality interval was selected by measuring the forward energy with the PSD. The measured pion pairs were grouped into four (Be+Be) and eight (Ar+Sc) average transverse momentum bins ranging from 0 to 600 MeV/c and from 0 to 450 MeV/c, respectively.

At vanishing relative momentum, the correlation function of Eq{.}~\ref{eq:corrfunc} reaches the intercept value $1 + \lambda$. As zero relative momentum is not accessible in the measurements, extrapolation from the region where two tracks are experimentally resolved, is needed. It is commonly observed that the intercept parameter $\lambda$ is less than 1, as also expected from Eq{.}~\ref{eq:lambda}. Based on this, one can interpret the value of this parameter.

\begin{figure}
\centering
\includegraphics[width=0.49\textwidth]{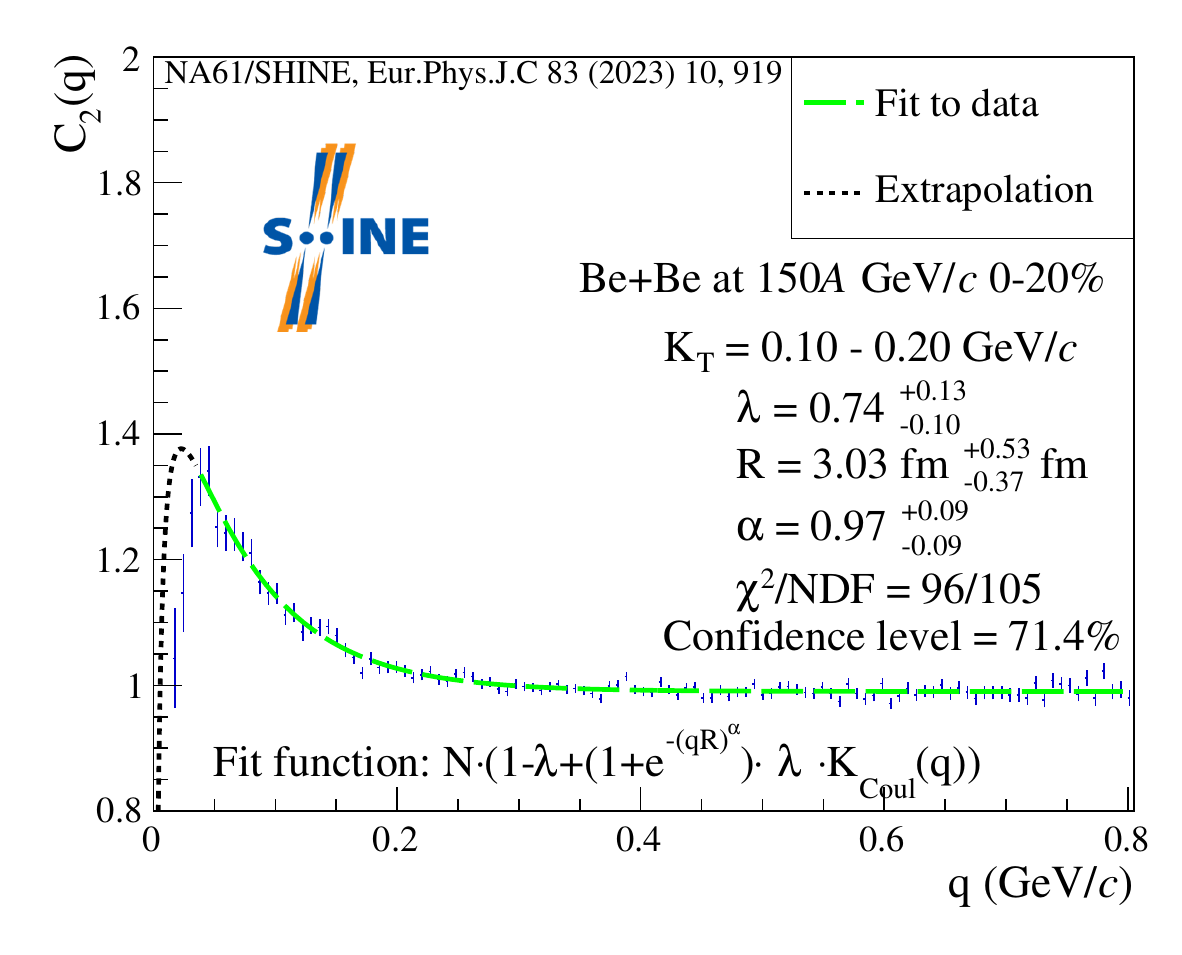}
\includegraphics[width=0.49\textwidth]{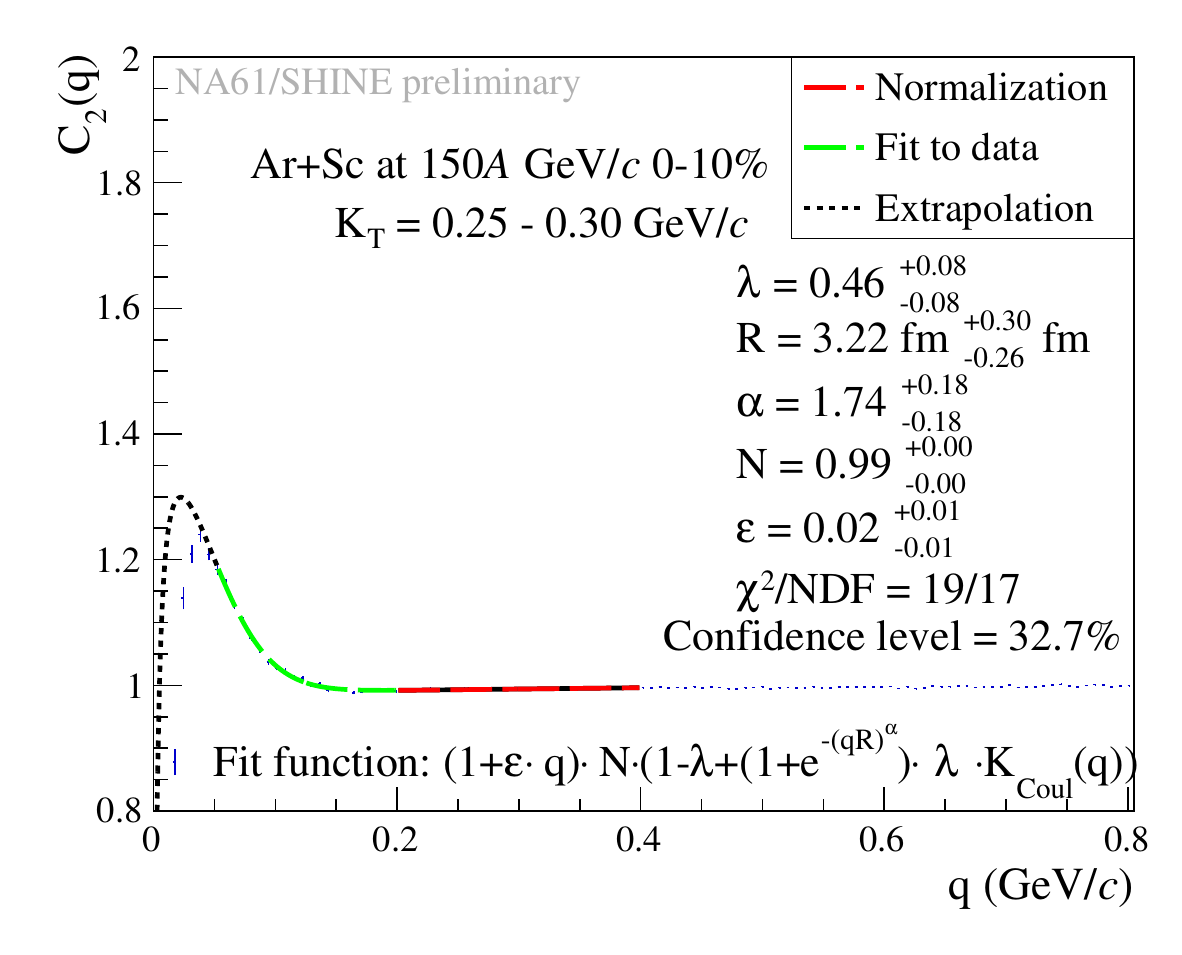}
\caption{Example fit of $\left(\pi^++\pi^+\right) + \left(\pi^-+\pi^-\right)$ femtoscopic correlation function for Be+Be and Ar+Sc, respectively.  Blue points with error bars represent the data, the green dashed line shows the fitted function with Coulomb correction. In the low-$q$ region, the black dotted line indicates the extrapolated function outside of the fit range, and red dashed line represent normalization to background for the Ar+Sc.}
\label{fig:examplefit}
\end{figure}

In this analysis, we investigate with like-charged particles that are influenced by Coulomb repulsion. The final state Coulomb effect has been neglected in the previously defined correlation function. The correction necessitated by this effect can be done by simply taking the ratio of $C^{\rm{Coulomb}}_2(q)$ and $C^0_2(q)$:
\begin{equation}\label{eq:coulombdef}
K_{\rm{Coulomb}}(q) = \frac{C^{\rm{Coulomb}}_2(q)}{C^0_2(q)}
\end{equation}
where $C^{\rm{Coulomb}}_2(q)$ is the interference of solutions of the two-particle Schrödinger equation; with a Coulomb-potential~\cite{Kincses:2019rug,Csanad:2019lkp}. The numerator in Eq{.}~\ref{eq:coulombdef} cannot be calculated analytically and requires numerical calculations.
A new treatment is required due to our assumption of L\'evy-shaped sources. We are utilizing a new method in our analysis for estimating the effect of Coulomb repulsion. The treatment includes the numerical calculation presented in Refs. ~\cite{Csanad:2019lkp, Csanad:2019cns} and the parametrization of its results.
Then, one can modify the correlation function, Eq{.}~\ref{eq:corrfunc}, to take the effect of the halo into account, by utilizing the Bowler-Sinyukov method~\cite{Sinyukov:1998fc, Bowler:1991vx}. The halo part contributes at very small values of relative momenta, \textit{q}. Therefore, it does not affect the source radii of the core part~\cite{Maj:2009ue}. Then the correlation function takes the following form:
\begin{equation}\label{eq:fitfunc}
C_2(q) = N\cdot \left( 1 + \epsilon \cdot q\right) \cdot \left(1 - \lambda + \lambda \cdot \left(1 + e^{-|qR|^\alpha} \right) \cdot K_{\rm{Coulomb}}(q) \right),
\end{equation}
where $N$ is introduced as normalization parameter and $K_{\rm{Coulomb}}(q)$ denotes the Coulomb correction. In the case of Ar+Sc collisions, an additional parameter ($\varepsilon$) responsible for describing the linearity of the background in the form of $\left(1 + \varepsilon \cdot q \right)$.
It is important to highlight that the Coulomb correction is calculated in the pair-center-of-mass (PCMS) system, while the measurement is often done in the longitudinally co-moving system (LCMS). The assumption of Coulomb correction in the one-dimensional HBT in LCMS picture is that the shape of the source is spherical, i.e.  $R_{\rm{out}} = R_{\rm{side}} = R_{\rm{long}} = R \equiv R_{\rm{LCMS}}$. The shape of the source, however, is spherical in the LCMS and not in the PCMS. Therefore, an approximate one-dimensional PCMS size parameter is needed.  In a recent study detailed in~\cite{Kurgyis:2020vbz}, the average PCMS radius was calculated using the following formula:  
\begin{equation}
\bar{R}_{\rm{PCMS}} = \sqrt{\frac{1 - \frac{2}{3}\beta_{\rm{T}^2}
}{1-\beta^2_{\rm{T}}}} \cdot R,
\end{equation} 
where $\beta_{\rm{T}} =  \frac{K_{\rm{T}}}{m_{\rm{T}}}$. To illustrate, an example fit using Eq{.}~\ref{eq:fitfunc} for Be+Be and Ar+Sc shown on Fig.~\ref{fig:examplefit}.

\section{Results}
\begin{figure}
\centering
\includegraphics[width=0.49\textwidth]{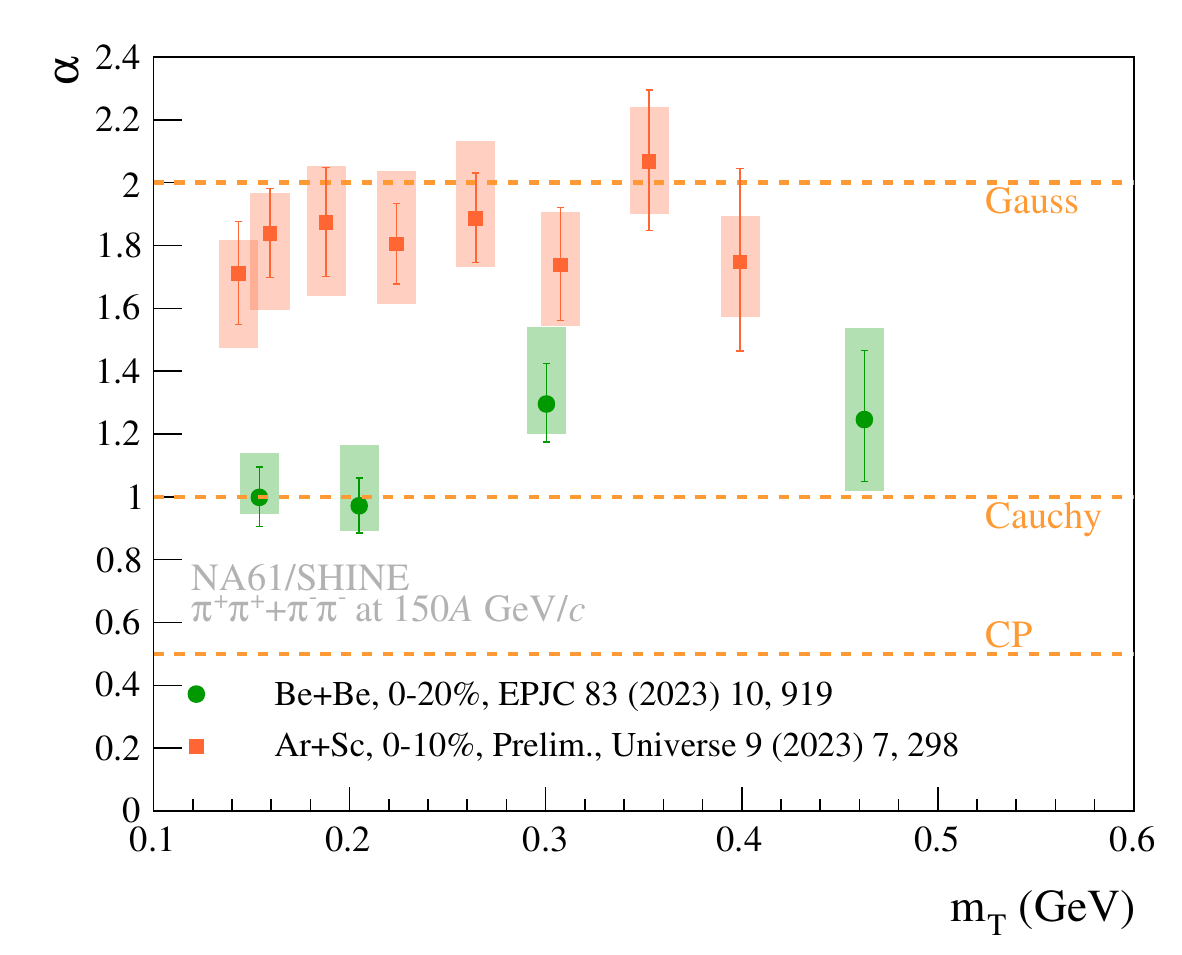}
\includegraphics[width=0.49\textwidth]{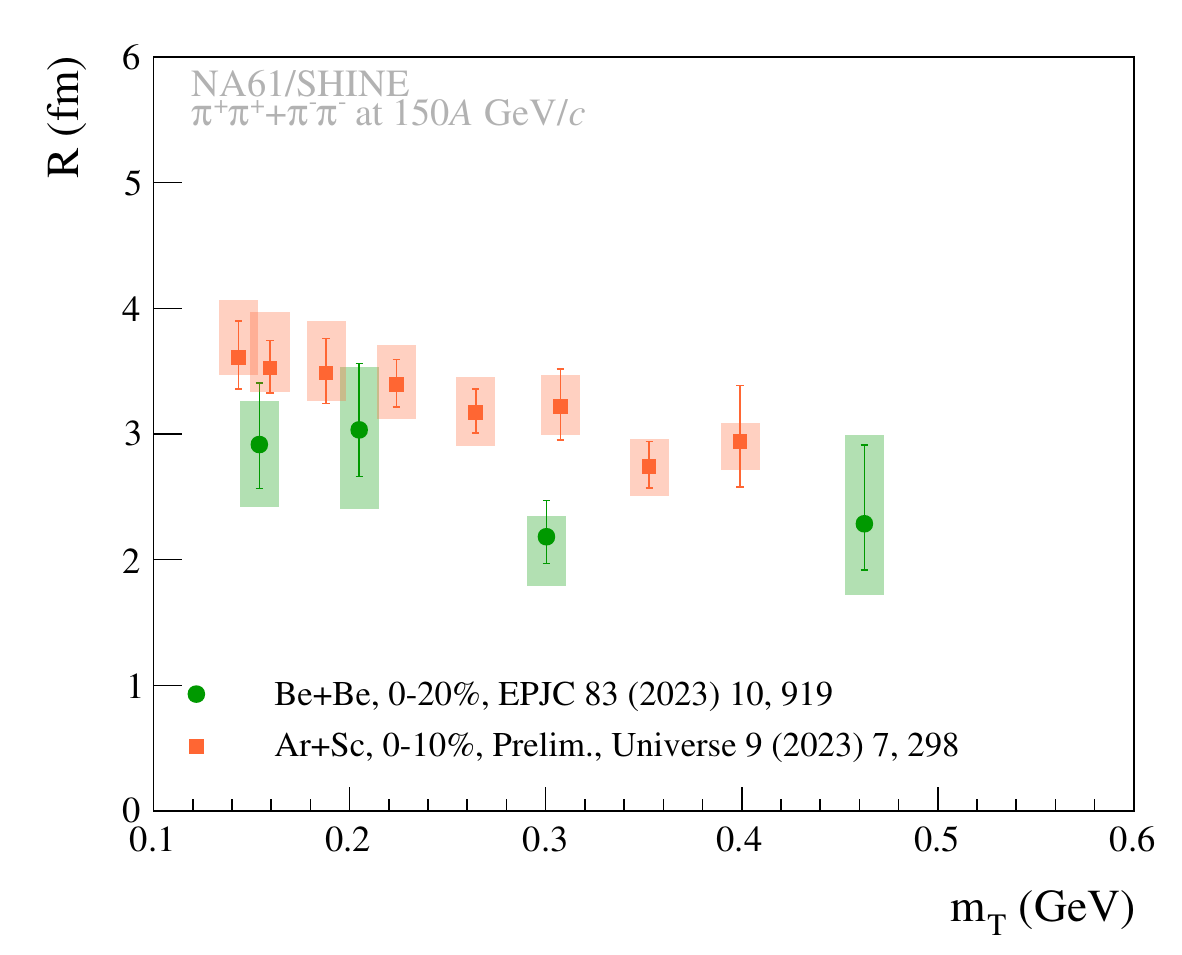}
\includegraphics[width=0.49\textwidth]{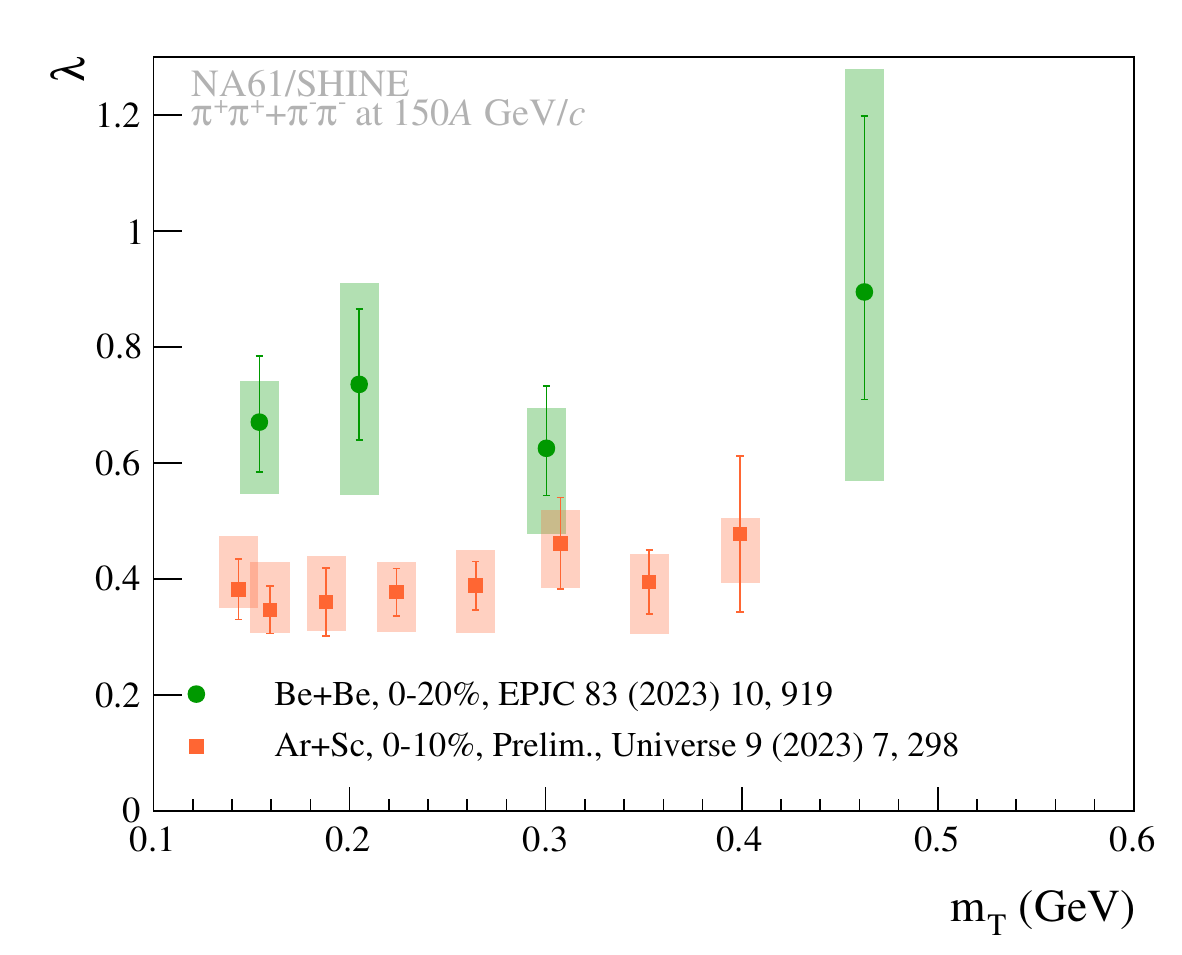}
\caption{The fit parameters, for $0$--$20$ \% central Be+Be at 150\textit{A} GeV/\textit{c} and $0$--$10$\% central Ar+Sc at 150\textit{A} GeV/\textit{c}, as a function of $m_{\rm{T}}$. Boxes denote systematic uncertainties, bars represent statistical uncertainties.}
\label{fig:results}
\end{figure}

The three physical parameters ($\alpha$, $R$, and $\lambda$) were measured in eight bins of pair transverse momentum,  $K_{\rm{T}}$. The three mentioned parameters were obtained through fitting the measured correlation functions with the formula shown in Eq{.}~\ref{eq:fitfunc}. The results were investigated regarding their transverse momentum dependence. In the following, we discuss the transverse $q$ mass dependence of $\alpha$, $R$, and $\lambda$; where transverse mass is expressed as $m_{\rm{T}} = \sqrt{m^2c^4+K_{\rm{T}}^2c^2}$.
As explained above, the shape of the source is often assumed to be Gaussian. The L\'evy stability exponent, $\alpha$, can be used to extract the shape of the tail of the source. Our results, shown in Fig. ~\ref{fig:results}, yield values for $\alpha$ between 1.5 and 2.0, which imply a source closer to the Gaussian shape than the one in Be+Be collisions~\cite{NA61SHINE:2023qzr}, but are often somewhat lower than the $\alpha = 2$ (Gaussian) case. The observed $\alpha$ parameter is also significantly higher than the conjectured value at the CP ($\alpha = 0.5$).  Based on the goodness-of-fit of the measured correlation functions, the L\'evy-stable source assumption is statistically acceptable. The L\'evy scale parameter $R$ is related to the length of homogeneity~\cite{Sinyukov1995} of the pion emitting source, albeit the concept of homogeneity length was worked out for Gaussian sources.  From simple hydrodynamical models ~\cite{Csorgo:1995bi, Csanad:2009wc} one obtains a decreasing trend of $R$ with transverse-mass.  In Fig. ~\ref{fig:results} a slight decrease for higher $m_{\rm{T}}$ values may be observed for both systems. This might be the result of transverse flow. Furthermore, a clear difference of parameter values is visible which might be attributed to the physical size differences of two collision systems. We observe an $R(m_{\rm{T}})$ trend not incompatible with $R \sim 1/\sqrt{m_{\rm{T}}}$ prediction, which is particularly interesting as this type of dependence was predicted for Gaussian sources ($\alpha = 2$)~\cite{Sinyukov:1994vg}. The origin of this $m_{\rm{T}}$ dependence is presently unknown, and it may form in the QGP or at a later stage. This phenomenon was also observed at RHIC~\cite{PHENIX:2017ino} and in simulations at RHIC and LHC energies~\cite{Kincses:2022eqq,Korodi:2022ohn}.
The final parameter to investigate is the correlation strength (also known as the intercept) parameter $\lambda$, defined in Eq{.}~\ref{eq:lambda}. The transverse mass dependence of $\lambda$ is shown in Fig. ~\ref{fig:results}. One may observe that this parameter is slightly dependent on $m_{\rm{T}}$ but mostly constant in the investigated range. When compared to measurements from RHIC Au+Au collisions~\cite{PHENIX:2017ino, Vertesi:2009wf, STAR:2009fks} and from SPS Pb+Pb interactions~\cite{Beker:1994qv, NA49:2007fqa}, an interesting phenomenon is observed. In the case of SPS experiments, there is no visible decrease at lower $m_{\rm{T}}$ values, but in the case of RHIC experiments, this appears. This decrease (or ``hole'') was interpreted in Ref.~\cite{PHENIX:2017ino} to be a sign of in-medium mass modification of $\eta'$. Our results, at the given statistical precision, do not indicate the presence of such a low-$m_{\rm{T}}$  hole. Furthermore, it is important to note that our values for $\lambda$ are significantly smaller than unity, which might imply that a significant fraction of pions are the decay products of long lived resonances or weak decays.

\section{Conclusion}

In this paper we presented the NA61/SHINE measurement of one-dimensional,  two-pion, femtoscopic correlation functions in 0-20\% centrality Be+Be collisions at 150A GeV/c and 0-10\% centrality Ar+Sc collisions at 150A GeV/c. We discussed the transverse mass dependence of the L\'evy source parameters. Results on the L\'evy scale parameter, $\alpha$, showed a deviation from Gaussian sources and are not in the vicinity of the conjectured value at the critical point. The L\'evy scale parameter, $R$, shows a visible decrease with $m_{\rm{T}}$. The correlation strength parameter, $\lambda$, does not show any significant $m_{\rm{T}}$ dependence,  in contrast to RHIC results, but similarly to earlier SPS measurements. In subsequent analyses, we also plan to measure Bose-Einstein correlations in larger systems, as well as at smaller energies, to continue exploring the phase diagram of the strongly interacting matter.

\acknowledgments
The author acknowledges support of the DKOP-23 Doctoral Excellence Program of the Ministry for Culture and Innovation, and was furthermore supported by K-138136 and K-138152 grants of the National Research, Development and Innovation Fund. 

\bibliography{Porfy__LevyHBT_Proceedings_WPCF23.bib}
\bibliographystyle{varenna.bst}

\end{document}